\documentclass[pra,showpacs,superscriptaddress]{revtex4}
\usepackage{graphicx,amsmath,bm,pifont}

\begin{document}

\title{Polarization-sensitive quantum-optical coherence tomography}

\author{Mark~C.~Booth}
\affiliation{Quantum Imaging Laboratory, Department of Biomedical
Engineering, Boston University, 8 Saint Mary's Street, Boston,
Massachusetts 02215}

\author{Giovanni~Di~Giuseppe}
\altaffiliation[Also at~]{Istituto Elettrotecnico Nazionale {\it
G.~Ferraris}, Strada delle Cacce 91, I-10153 Torino, Italy.}
\affiliation{Quantum Imaging Laboratory, Department of Electrical
\& Computer Engineering, Boston University, 8 Saint Mary's Street,
Boston, Massachusetts 02215}

\author{Bahaa~E.~A.~Saleh}
\affiliation{Quantum Imaging Laboratory, Department of Electrical
\& Computer Engineering, Boston University, 8 Saint Mary's Street,
Boston, Massachusetts 02215}

\author{Alexander~V.~Sergienko}
\affiliation{Quantum Imaging Laboratory, Department of Electrical
\& Computer Engineering, Boston University, 8 Saint Mary's Street,
Boston, Massachusetts 02215}\affiliation{Quantum Imaging
Laboratory, Department of Physics, Boston University, 8 Saint
Mary's Street, Boston, Massachusetts 02215}

\author{Malvin~C.~Teich}
\email{teich@bu.edu} \homepage[Quantum Imaging Laboratory
homepage:~]{http://www.bu.edu/qil} \affiliation{Quantum Imaging
Laboratory, Department of Biomedical Engineering, Boston
University, 8 Saint Mary's Street, Boston, Massachusetts
02215}\affiliation{Quantum Imaging Laboratory, Department of
Electrical \& Computer Engineering, Boston University, 8 Saint
Mary's Street, Boston, Massachusetts 02215}\affiliation{Quantum
Imaging Laboratory, Department of Physics, Boston University, 8
Saint Mary's Street, Boston, Massachusetts 02215}

\date{\today}

\renewcommand{\baselinestretch}{2}\small\normalsize
\begin{abstract}
We set forth a polarization-sensitive quantum-optical coherence
tomography (PS-QOCT) technique that provides axial optical
sectioning with polarization-sensitive capabilities.  The
technique provides a means for determining information about the
optical path length between isotropic reflecting surfaces, the
relative magnitude of the reflectance from each interface, the
birefringence of the interstitial material, and the orientation of
the optical axis of the sample. PS-QOCT is immune to sample
dispersion and therefore permits measurements to be made at depths
greater than those accessible via ordinary optical coherence
tomography. We also provide a general Jones matrix theory for
analyzing PS-QOCT systems and outline an experimental procedure
for carrying out such measurements.
\end{abstract}


\maketitle

\section{INTRODUCTION}
Optical coherence tomography (OCT) has become a well-established
imaging technique~\cite{Huang91,Fujimoto95,Fercher96,Schmidt99}
with applications in ophthalmology~\cite{Hee95}, intravascular
measurements~\cite{Brezinski99_IEEE,Brezinski99}, and
dermatology~\cite{Welzel01}. It is a form of range-finding that
makes use of the second-order coherence properties of a classical
optical source to effectively section a reflective sample with a
resolution governed by the coherence length of the source. OCT
therefore makes use of sources of short coherence length (and
consequently broad spectrum), such as superluminescent diodes
(SLDs) and ultrashort-pulsed lasers.  As broad bandwidth sources
are developed to improve the resolution of OCT techniques,
material dispersion has become more pronounced. The deleterious
effects of dispersion broadening limits the achievable resolution
as has been recently emphasized~\cite{Drexler01}.

To further improve the sensitivity of OCT, techniques for handling
dispersion must be implemented. In the particular case of
ophthalmologic imaging, one of the most important applications of
OCT, the retinal structure is located behind a comparatively large
body of dispersive ocular media~\cite{Hitzenberger99}. Dispersion
increases the width of the coherence envelope of the probe beam
and results in a reduction in axial resolution and fringe
visibility~\cite{Hitzenberger02}. Current techniques for
depth-dependent dispersion compensation include the use of
dispersion-compensating elements in the optical
set-up~\cite{Hitzenberger99,Smith02} or employ {\it a posteriori}
numerical methods~\cite{Fercher01,Fercher02}.  For these
techniques to work, however, the object dispersion must be known
and well characterized so that the appropriate optical element or
numerical algorithm can be implemented.

Over the past several decades, a number of non-classical (quantum)
sources of light have been developed~\cite{Teich88,Teich90} and it
is natural to inquire whether making use of any of these sources
might be advantageous for tomographic imaging.  An example of such
a nonclassical source is spontaneous parametric down-conversion
(SPDC)~\cite{Klyshko67,Harris67,Giallorenzi68,Kleinman68,Burnham70,Larchuk95},
a nonlinear process that produces entangled beams of light. This
source, which is broadband, has been utilized to demonstrate a
number of interference effects that cannot be observed using
traditional classical sources of light.  We make use of this
unique feature in quantum-optical coherence tomography
(QOCT)~\cite{Abouraddy02_PRA}, where fourth-order interference is
used to provide range measurements analogous to those currently
obtained using classical OCT, but with the added advantage of
even-order dispersion
cancellation~\cite{Franson92,Steinberg92_PRL,Larchuk95_PRA}. We
have recently demonstrated the dispersion immunity of these
tomographic measurements in comparison to standard optical
coherence tomography techniques~\cite{Nasr03}.

In this paper, we present a method for polarization-sensitive QOCT
(PS-QOCT) measurements, where one can detect a change in the
polarization state of light reflected from a layered
sample~\cite{Hee92}. This state change arises from scattering and
birefringence in the sample and is enhanced in specimens that have
an organized linear structure.  Tissue that contains a high
content of collagen or other elastin fibers, such as tendons,
muscle, nerve, or cartilage, are particularly suited to
polarization-sensitive measurements~\cite{deBoer99}. A variation
in birefringence can be indicative of a change in functionality,
integrity, or viability of biological tissue.

\section{GENERAL MATRIX THEORY FOR PS-QOCT}
\begin{figure}
\includegraphics[clip,width=14cm]{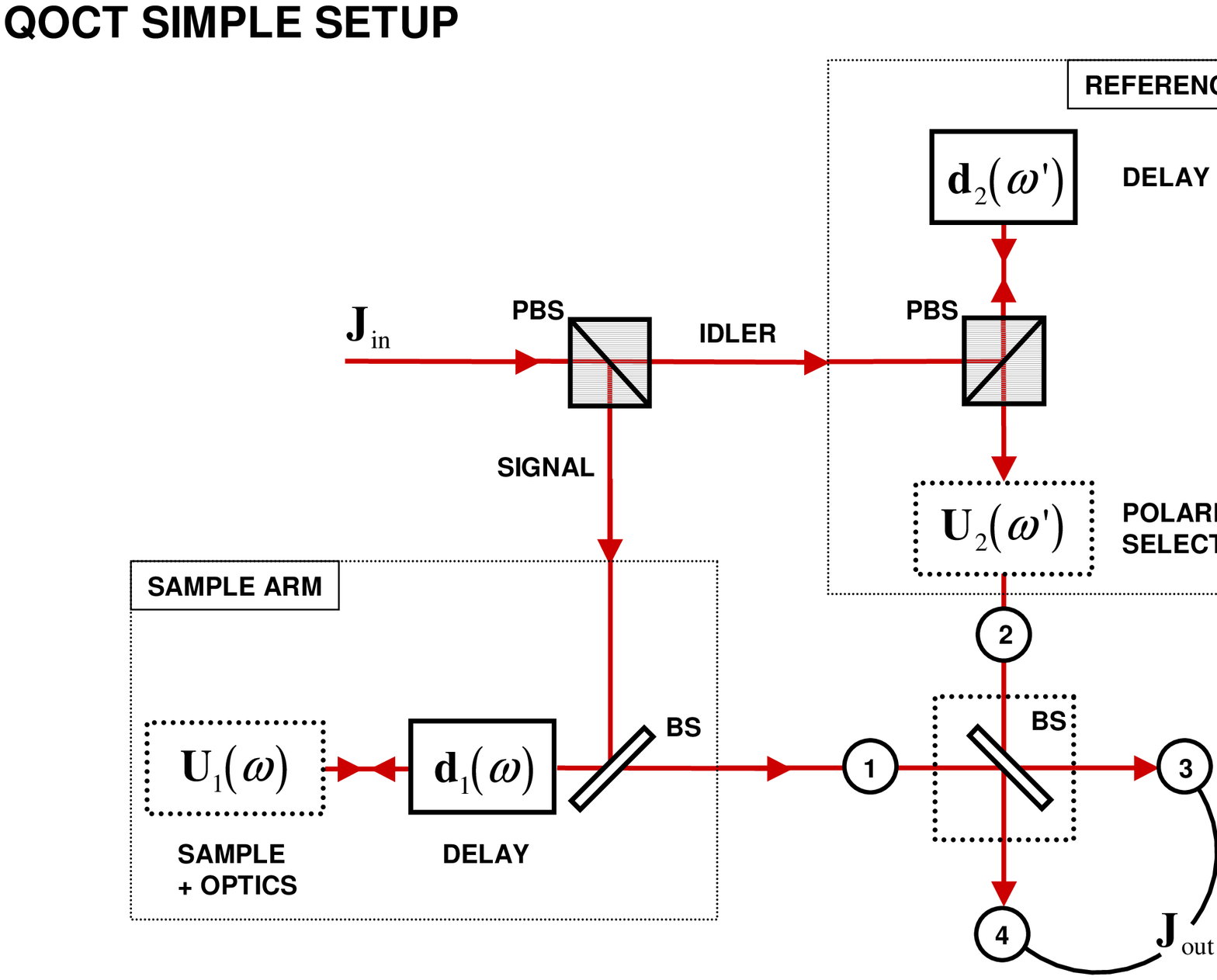}
\caption{Conceptual diagram of polarization-sensitive
quantum-optical coherence tomography (PS-QOCT). The system is
based on a Mach-Zehnder interferometer in which twin photons from
SPDC, represented by the vector $\rm{{\bf J}_{in}}$, are separated
into two arms at a polarizing beam splitter (PBS).  The signal
photon at angular frequency $\omega$ travels in the sample arm and
experiences a path delay $\rm{\bf d}_1$ as well as an arbitrary
polarization rotation described by $\rm{\bf U}_1$.  The reference
arm contains the idler photon at angular frequency $\omega'$ which
experiences a path delay $\rm{\bf d}_2$ and a user-selected
polarization rotation $\rm{\bf U}_2$.  Paths \ding{192} and
\ding{193} impinge on a final beam splitter (BS) which mixes the
spatial/polarization modes into paths \ding{194} and \ding{195}.
$\rm{{\bf J}_{out}}$ represents the final twin-photon Jones vector
from which the fields at the detectors and the final coincidence
rate are computed.} \label{fig:QOCT-theory-setup}
\end{figure}

We present the theory for PS-QOCT according to the simplified
diagram for an experimental setup given in
Fig.~\ref{fig:QOCT-theory-setup}. Using a Jones-matrix formalism
similar to that in Ref.~\cite{Abouraddy02_JOSAB2}, we start by
defining a twin-photon Jones vector $\rm{{\bf J}_{in}}$
\begin{equation}
    \rm{{\bf J}_{in}} =
   \begin{bmatrix}
     {\hat a}_{\rm s}(\omega) \, {\bf e}_{\rm s}\\
     {\hat a}_{\rm i}(\omega')\, {\bf e}_{\rm i}
  \end{bmatrix},
  \label{eq:Jin}
\end{equation}
where ${\hat a}_{\rm s}(\omega)$ and ${\hat a}_{\rm i}(\omega')$
are the annihilation operators for the signal-frequency mode
$\omega$ and the idler-frequency mode $\omega'$, respectively. The
vectorial polarization information for the signal and idler field
modes are contained in ${\bf e}_j$, ($j = {\rm s, \, i}$). For
example, if we utilize collinear type-II SPDC from a second-order
nonlinear crystal (NLC) to generate a pair of orthogonally
polarized photons, the ${\bf e}_j$ reduce to the familiar Jones
vectors:
\begin{eqnarray}
  {\bf e}_{\rm s} &=& \genfrac{[}{]}{0pt}{}{1}{0} \quad\quad {\rm
  (vertical)}\nonumber\\
  {\bf e}_{\rm i} &=& \genfrac{[}{]}{0pt}{}{0}{1} \quad\quad {\rm (horizontal)}
    \label{eq:sig-id-pol-vecs}
\end{eqnarray}
for signal and idler, respectively.

The twin photons collinearly impinge on the input port of a
polarizing beam splitter (PBS) from which the signal photon is
reflected into the sample arm and the idler photon is transmitted
into the reference arm. We assume that the polarization in each
arm is independent of that in the other arm until the final beam
splitter. We also assume that all optical elements within the
interferometer are linear and deterministic.

The delay accumulated by the signal and idler beams in each path
is represented by the $2 \times 2$ matrix
\begin{equation}
    \rm{\bf D} =
  \begin{bmatrix}
    \rm{\bf d}_1(\omega) & 0 \\
    0 & \rm{\bf d}_2(\omega')
  \end{bmatrix},
  \label{eq:delay}
\end{equation}
where $\rm{\bf d_1}(\omega)$ and $\rm{\bf d_2}(\omega')$ represent
the Jones matrices that describe the delay for the sample and
reference arms, respectively. The polarization state in each arm
is represented by the matrix
\begin{equation}
    \rm{\bf U} =
  \begin{bmatrix}
    \rm{\bf U}_1(\omega) & 0 \\
    0 & \rm{\bf U}_2(\omega')
  \end{bmatrix}.
  \label{eq:polarization}
\end{equation}
In the experimental realization of PS-QOCT, the matrix $\rm{\bf
U}_1(\omega)$ represents the properties of the sample {\it plus}
any other polarization elements in the sample arm, whereas
$\rm{\bf U}_2(\omega')$ represents the user-selected polarization
state in the reference arm.

The mixing of the polarizations from each path, which occurs at
the final beam splitter, is represented by the transformation
matrix
\begin{equation}
   \rm{{\bf T}_{BS}} =
  \begin{bmatrix}
    \rm{\bf T}_{31} & \rm{\bf T}_{32} \\
    \rm{\bf T}_{41} & \rm{\bf T}_{42}
  \end{bmatrix},
  \label{eq:BS}
\end{equation}
where each element ${\rm \bf T}_{kl}$ ($k =$~3, 4 and $l =$~1, 2)
is a $2 \times 2$ matrix that represents the mixing of independent
polarization modes from the input paths \ding{192} and \ding{193}
prior to detection in paths \ding{194} and \ding{195}. For
example, $\rm{\bf T}_{31}$ is the Jones matrix that represents the
transformation of input spatial mode~1 into output spatial mode~3.

The Jones vector $\rm{ {\bf J}_{out}}$ that describes the field
operators at the output of the final beam splitter, can be
computed from the product of the previously defined matrices in
Eqs.~(\ref{eq:delay})~--~(\ref{eq:BS}) as
\begin{equation}
    \rm{{\bf J}_{out}} =
    \rm{{\bf T}_{BS}} \, \rm{\bf U} \, \rm{\bf D} \, \rm{{\bf
    J}_{in}} =
    \begin{bmatrix}
    \rm{\bf T}_{31}\rm{\bf U}_1\rm{\bf d}_1 & \rm{\bf T}_{32}\rm{\bf U}_2\rm{\bf d}_2 \\
    \rm{\bf T}_{41}\rm{\bf U}_1\rm{\bf d}_1 & \rm{\bf T}_{42}\rm{\bf U}_2\rm{\bf d}_2
  \end{bmatrix}
  \, \rm{{\bf J}_{in}}.
  \label{eq:Jout}
\end{equation}
>From this equation, the fields in paths \ding{194} and \ding{195}
arriving at each of the two detectors can be written in the time
domain as
\begin{equation}
    {\hat {\bf E}}^{(+)}_{3}(t_3) = \int {\rm d}\omega \, {\rm e}^{-{\rm i} \omega t_3}
    \, {\hat a}_{\rm s}(\omega) \, {\bf e}_{3 {\rm s}}
    + \int {\rm d}\omega' \, {\rm e}^{-{\rm i} \omega' t_3}
    \, {\hat a}_{\rm i}(\omega') \, {\bf e}_{3 {\rm i}}
  \label{eq:E3}
\end{equation}
\begin{equation}
    {\hat {\bf E}}^{(+)}_{4}(t_4) = \int {\rm d}\omega \, {\rm e}^{-{\rm i} \omega t_4}
    \, {\hat a}_{\rm s}(\omega) \, {\bf e}_{4 {\rm s}}
    + \int {\rm d}\omega' \, {\rm e}^{-{\rm i} \omega' t_4}
    \, {\hat a}_{\rm i}(\omega') \, {\bf e}_{4 {\rm i}},
  \label{eq:E4}
\end{equation}
where
\begin{eqnarray}
    {\bf e}_{3{\rm s}} = {\bf T}_{31}\,{\bf U}_{1}\,{\bf d}_{1}\,{\bf e}_{\rm s}
                    \hspace{3cm}
    {\bf e}_{3{\rm i}} = {\bf T}_{32}\,{\bf U}_{2}\,{\bf d}_{2}\,{\bf e}_{\rm i}
                    \nonumber \\
    {\bf e}_{4{\rm s}} = {\bf T}_{41}\,{\bf U}_{1}\,{\bf d}_{1}\,{\bf e}_{\rm s}
                    \hspace{3cm}
    {\bf e}_{4{\rm i}} = {\bf T}_{42}\,{\bf U}_{2}\,{\bf d}_{2}\,{\bf e}_{\rm i}
    \label{eq:es_ei}
\end{eqnarray}
describes each of the transformations of the signal and idler
polarizations that contribute to the final fields in \ding{194}
and \ding{195} at the detectors.

>From the field at each of the detectors, the two-photon amplitude
can be written as
\begin{equation}
    A_{jk}(t_3,t_4) = \langle0|{\hat E}^{(+)}_{3j}(t_3){\hat
    E}^{(+)}_{4k}(t_4)|\Psi\rangle,
  \label{eq:amplitude}
\end{equation}
where $j$ and $k$ represent two orthogonal polarization bases such
as horizontal/vertical (H/V), right/left circular (R/L), or
(45$^{\circ}$/-45$^{\circ}$). The ket $|\Psi\rangle$ represents
the two-photon state at the output of the nonlinear crystal,
defined as
\begin{equation}
    |\Psi\rangle = \int {\rm d}\Omega  \,  \Phi(\Omega) \,
    \hat{a}^\dag_{\rm s}(\omega_0 + \Omega) \, \hat{a}^\dag_{\rm i}(\omega_0 - \Omega)
    |0\rangle,
  \label{eq:two_photon_state}
\end{equation}
where $\Phi(\Omega)$ is the state function~\cite{Giuseppe02} that
governs the spatio-temporal properties of the signal and idler
photons at angular frequency $\omega_0 \pm \Omega$.  The state
function is given by $\Phi(\Omega) = L\,{\rm sinc}[\Delta
k_{z}(\Omega)L/2]$, where $L$ is the crystal length and $\Delta
k_{z}(\Omega)$ is the wave-vector mismatch in the $z$- or
phase-matching direction. If the state function $\Phi(\Omega)$ is
symmetric about the center frequency or $\Phi(-\Omega) =
\Phi(\Omega)$, the SPDC spectrum becomes $|\Phi(\Omega)|^2$.

We assume that the detection apparatus is slow and independent of
polarization so that the final coincidence rate $R$ is computed as
the magnitude-square of the two-photon amplitude summed over each
polarization mode, integrated over time:
\begin{equation}
    R = \int dt_3 \int dt_4\sum_{j}\sum_{k}|A_{jk}(t_3,t_4)|^{2}.
  \label{eq:rate}
\end{equation}

\section{SIMPLIFIED CONFIGURATION FOR PS-QOCT}
\begin{figure}
\includegraphics[clip,width=14cm]{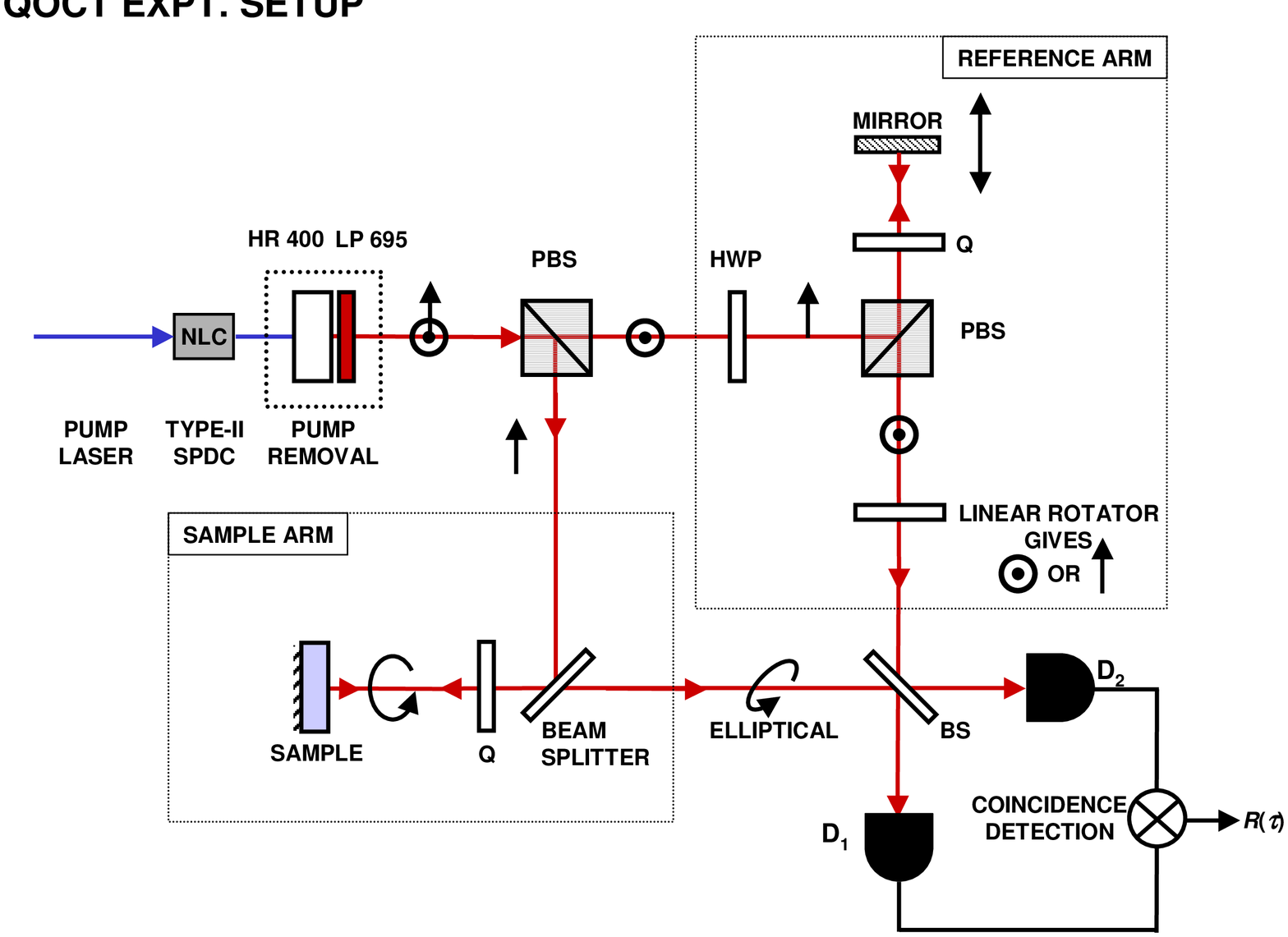}
\caption{Possible implementation of polarization-sensitive
quantum-optical coherence tomography (PS-QOCT).  A narrow-band
pump laser at a wavelength of 400~nm pumps a 1.5-mm-thick
$\beta$-barium borate (BBO) nonlinear crystal (NLC) oriented for
type-II, collinear SPDC with a center wavelength of 800~nm. The
pump beam is removed from the SDPC by use of a highly reflective
mirror (HR~400) centered at the pump wavelength concatenated with
a long pass filter (LP~695). The vertical and horizontal
components in the SPDC beam are separated by a polarizing beam
splitter (PBS) into the reference arm and sample arm of a
Mach-Zehnder interferometer.  The reference arm consists of a
variable path-length delay comprised of a half-wave plate (HWP), a
second polarizing beam splitter (PBS), a quarter-wave plate (Q),
and a translational mirror. The final polarization of the
reference beam (indicated as $\odot$) can be oriented to either
vertical or horizontal by a linear rotator prior to the final beam
splitter (BS). The sample arm consists of a beam splitter and a
quarter-wave plate (Q) so that circularly polarized light is
normally incident on the sample. The back-reflected light from the
sample, which in general has elliptical polarization, mixes with
the delayed reference beam at the final beam splitter (BS). The
outputs from the BS are directed to two single-photon counting
detectors. The coincidence rate $R(\tau)$ for photons arriving at
the two detectors, as a function of the path-length delay $c\tau$,
are recorded in a time window determined by a coincidence-counting
detection circuit (indicated as $\otimes$).}
\label{fig:QOCT-setup}
\end{figure}

It is now useful to consider a specific experimental configuration
from which we can define expressions for the Jones matrices in
Eqs.~(\ref{eq:delay})~--~(\ref{eq:BS}) and calculate the quantum
interferogram. One particular experimental configuration for
PS-QOCT is based on the Mach-Zehnder interferometer and is shown
in Fig.~\ref{fig:QOCT-setup}.

The elements of the Jones matrix in Eq.~(\ref{eq:delay}),
representing the delay in each path, are given simply by ${\bf
d}_1(\omega) = {\bf I} \, {\rm e}^{{\rm i} \omega z_{\rm s}/c}$
and ${\bf d}_2(\omega') = {\bf I} \, {\rm e}^{{\rm i} \omega'
z_{\rm r}/c}$, where ${\bf I}$ is the identity matrix, $c$ is the
speed of light in the medium, and $z_{\rm s}$, $z_{\rm r}$ are the
path lengths in the sample and reference arms, respectively. We
define a path-delay difference $\tau = (z_{\rm r} - z_{\rm s})/c$
between the reference and sample arms that becomes our
experimental parameter in the final expression for the measured
coincidence rate $R(\tau)$. We assume that the final beam splitter
faithfully transmits and reflects each input polarization mode.
The elements of ${\bf T}_{\rm BS}$ in Eq.~(\ref{eq:BS}) thus
become ${\bf T}_{31} = {\bf T}^{\dag}_{42} = t\,{\bf I}$ and ${\bf
T}_{32} = - {\bf T}^{\dag}_{41} = r^{\ast}\,{\bf I}$, where $\dag$
designates a matrix transpose and conjugation, and $t$ and $r$
represent the amplitude transmittance and reflectance of the beam
splitter, respectively.

Without having to specify the elements of ${\bf U}$ in
Eq.~(\ref{eq:polarization}), an expression for the measured
coincidence rate as a function of the path-delay difference $\tau$
is calculated to be
\begin{equation}
  R(\tau) \propto \Lambda_{0} - {\mathcal V}_{\rm BS}\,{\rm Re}\left[\Lambda(2\tau)\right],
    \label{eq:R-simple}
\end{equation}
where $\Lambda_{0}$ and $\Lambda(\tau)$ are defined as
\begin{equation}
  \Lambda_{0} = \int d\Omega\,|\Phi(\Omega)|^{2}\,
            [{\bf e}^{\dag}_{\rm s}\,{\bf U}^{\dag}_{1}(\omega_0 + \Omega)\,{\bf U}_{1}(\omega_0 + \Omega)
            \,{\bf e}_{\rm s}]\,
            ({\bf e}^{\dag}_{\rm i}\,{\bf U}^{\dag}_{2}\,{\bf U}_{2}\,{\bf e}_{\rm i})
    \label{eq:lambda_zero}
\end{equation}
and
\begin{equation}
  \Lambda({\tau}) = \int d\Omega\,|\Phi(\Omega)|^{2}\,
            F(\omega_0 + \Omega)\,F^{\ast}(\omega_0 - \Omega)\,
            {\rm e}^{-{\rm i}\Omega\tau},
    \label{eq:lambda_tau}
\end{equation}
representing constant and varying contributions to the quantum
interferogram, respectively. The function $F(\omega)$, which
includes all of the sample properties, is given by
\begin{equation}
  F(\omega) = {\bf e}^{\dag}_{\rm i}\,{\bf U}^{\dag}_{2}\,{\bf U}_{1}(\omega)\,{\bf e}_{\rm
  s}.
    \label{eq:F}
\end{equation}
The parameter ${\mathcal V}_{\rm BS} = 2(|r|^2|t|^2)/(|r|^4 +
|t|^4)$ in Eq.~(\ref{eq:R-simple}) represents a visibility factor
for a lossless beam splitter with arbitrary transmittance
(${\mathcal V}_{\rm BS} = 1$ when $|r|^2 = |t|^2 = 1/2$). We
assume that the optical elements in the reference arm are
frequency independent across the bandwidth of the light-source
spectrum. Equation~(\ref{eq:lambda_tau}) is a generalization of
Eq.~(8) in Ref.~\cite{Abouraddy02_PRA}, where the function $F$
contains polarization-dependent information about the sample.

It is clear from Eq.~(\ref{eq:lambda_tau}) that the sample is
simultaneously probed at two frequencies, $\omega_0 + \Omega$ and
$\omega_0 - \Omega$, and that for a frequency-entangled two-photon
state such that produced by SPDC, even-order dispersion from the
sample is cancelled in PS-QOCT. The effectiveness of even-order
dispersion cancellation is related to the spectrum of the source
used for SPDC.  Since we assume a cw-pump source in
Eq.~(\ref{eq:two_photon_state}), this leads to signal and idler
photons that are exactly frequency anti-correlated.  In this case,
even-order dispersion cancellation is perfect. As the bandwidth of
the SPDC pump source is increased, the requirements for exact
frequency anti-correlation are relaxed and dispersion cancellation
is degraded. It is apparent that the delay $\tau$ can be adjusted
to target specific regions in the sample from which polarization
information can be extracted by scanning the parameters of the
user-selected polarization rotator $\rm{\bf U}_2$. This
experimental method is similar to those used in quantum
ellipsometry, as discussed in Ref.~\cite{Abouraddy02_JOSAB2}.

In the following section, we consider a specific construct for
Eq.~(\ref{eq:polarization}) that defines the optics in the
experimental setup represented in Fig.~\ref{fig:QOCT-setup}. Once
we derive an expression valid for an arbitrary sample, we consider
several special cases in an effort to understand the nature of the
information contained in the quantum interferogram.

\section{ROLE OF POLARIZATION IN PS-QOCT}
To facilitate the description of PS-QOCT, we make use of the Pauli
spin matrices
\begin{equation}
\mbox{\boldmath$\sigma$}_1= \left[
    \begin{array}{ccc}
    0 & 1 \\
    1 & 0
    \end{array}\right]
            \hspace{2cm}
\mbox{\boldmath$\sigma$}_2= \left[
    \begin{array}{ccc}
    0 & -{\rm i} \\
    {\rm i} & 0
    \end{array}\right]
            \hspace{2cm}
\mbox{\boldmath$\sigma$}_3= \left[
    \begin{array}{ccc}
    1 & 0 \\
    0 & -1
    \end{array}\right],
    \label{eq:Pauli_matrices}
\end{equation}
from which any 2~x~2 Hermitian matrix can be defined as ${\rm \bf
A} = c_0\,{\rm \bf I} + {\bf c} \cdot {\bm \sigma}$, where ${\bf
c} \equiv (\mbox{$ c $}_1,\mbox{$ c $}_2,\mbox{$ c $}_3)$, ${\bm
\sigma} \equiv
(\mbox{\boldmath$\sigma$}_1,\mbox{\boldmath$\sigma$}_2,\mbox{\boldmath$\sigma$}_3)$,
and ${\bf c} \cdot {\bm \sigma}$ denotes the scalar product of
vectors ${\bf c}$ and ${\bm \sigma}$. We first define
Eq.~(\ref{eq:polarization}) for $N$ reflective layers, where each
reflection is assumed to be isotropic, then consider the special
cases of a single and double reflector.

\subsection{$N$ reflective layers}
\begin{figure}
\includegraphics[clip,width=14cm]{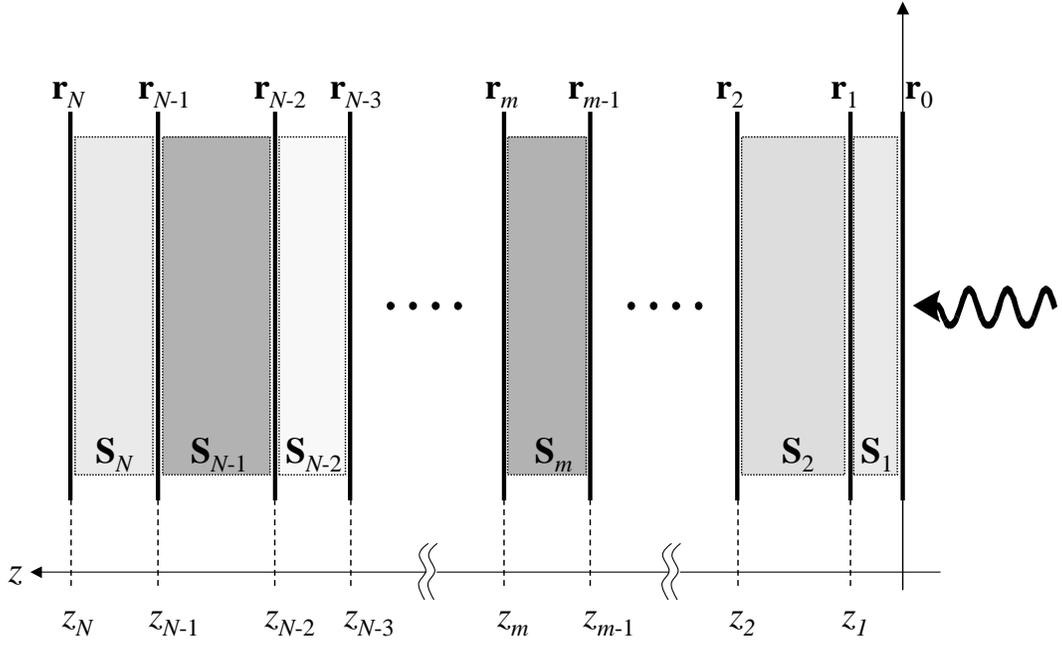}
\caption{Sample comprised of $N$ reflective layers.  The probe
beam is incident at the right. Each interface at position $z_m$ is
described by a reflectance matrix ${\rm{\bf r}}_m(\omega)$.  The
optical properties of the sample layers between interfaces are
described by the Jones matrix ${\rm{\bf S}}_m$.}
\label{fig:N-layer-sample}
\end{figure}

We begin with a sample comprised of $N$ reflective layers, each
with an interface defined by a reflectance matrix ${\rm \bf
r}_m(\omega)$, as shown in Fig.~\ref{fig:N-layer-sample}. The
material properties of each layer are represented by a Jones
matrix ${\rm \bf S}_m$ that is assumed to be deterministic.  The
Jones matrix ${\rm \bf S}_m$ is a product of: an average phase
delay; rotation matrices ${\rm \bf R}_m$ to account for the
orientation $\alpha_m$ of the fast axis of the sample with respect
to the horizontal axis; and the Jones matrix ${\bf b}_m$ for a
linear retarder with its fast axis oriented along the horizontal
axis~\cite{OCTHandbook}. If we ignore losses due to absorption,
then for a single layer of thickness $d_m = (z_m - z_{m-1})$, the
Jones matrix is given by
\begin{eqnarray}
  {\bf S}_m(d_m,\alpha_m,\omega) &=& {\rm e}^{{\rm i}\Delta_m(d_m,\omega)}\,
                {\bf R}_m(\alpha_m)\,{\bf b}_m(\delta_m)\,{\bf R}^{\dag}_m(\alpha_m)
                \nonumber\\
                             &\equiv& {\rm e}^{{\rm i}\Delta_m(d_m,\omega)}\,
                {\bf B}_m(d_m,\alpha_m,\omega)
    \label{eq:N-layer-sample}
\end{eqnarray}
where $\Delta_m(d_m,\omega) =  \omega\,{\bar n}\, d_m/c $ is the
average phase delay of the signal photon at angular frequency
$\omega$ attained by propagating through a layer with average
refractive index $\bar n = (n_{\rm o} + n_{\rm e})/2$. The
single-pass retardation in the layer is given by
$\delta_m(d_m,\omega) = \omega\,\Delta n \, d_m /c $ where $\Delta
n = n_{\rm o} - n_{\rm e}$ is the difference in refractive indices
along the fast and slow axes of the medium. The rotation matrix
${\rm \bf R}_m(\alpha_m)$ and the Jones matrix ${\bf
b}_m(\delta_m)$ for the linear retarder are given by
\begin{equation}
    {\bf R}_m(\alpha_m) = {\rm e}^{-{\rm i}\alpha_m\,\mbox{\boldmath$\sigma$}_2}
            \hspace{0.25cm} {\rm and} \hspace{0.25cm}
    {\bf b}_m(\delta_m) = {\rm e}^{{\rm
    i}(\delta_m/2)\,\mbox{\boldmath$\sigma$}_3},
  \label{eq:rotation_and_retarder}
\end{equation}
respectively, where in general ${\rm e}^{-{\rm
i}\gamma\,\mbox{\boldmath$\sigma$}} = (\cos\gamma) \,{\bf I} -
({\rm i}\sin\gamma) \, \mbox{\boldmath$\sigma$}$.

A complete transfer function describing the entire sample in
Fig.~\ref{fig:N-layer-sample} is therefore constructed as
\begin{eqnarray}
  {\bf H}(\omega) &=& \sum_{m=0}^{N}\,{\bf S}_{1}{\bf S}_{2}\cdot\cdot\cdot{\bf S}_{m-1}{\bf S}_{m}
                                    \,{\bf r}_{m}
                                  \,{\tilde{\bf S}_{m}}{\tilde{\bf S}}_{m-1}\cdot\cdot\cdot{\tilde{\bf S}}_{2}{\tilde{\bf S}}_{1}
              \nonumber \\
        &=& \sum_{m=0}^{N}\,{\rm e}^{{\rm i}2\varphi^{(m)}}
                          \,{\bf B}^{(m)}\,{\bf r}_{m}\,{\tilde{\bf B}^{(m)}},
    \label{eq:N-layer-function}
\end{eqnarray}
where $\varphi^{(0)} = 0$, $B^{(0)} = {\bf I}$, and the tilde
operation denotes a matrix that takes an argument at a negative
angle, viz. ${\tilde{\bf S}_{m}}(\alpha) = {\bf S}_{m}(-\alpha)$.
The accumulation of all phases up to interface $m$ is given by
\begin{equation}
  \varphi^{(m)} = \sum_{l=1}^{m}\Delta_l(d_l,\omega)
    \label{eq:sum-phases}
\end{equation}
and the accumulated effect of birefringence up to interface $m$ is
expressed via
\begin{equation}
  {\bf B}^{(m)} = \prod_{l=1}^{m}{\bf B}_l(d_l,\alpha_l,\omega).
    \label{eq:sum-bifi}
\end{equation}

Since the principal axes of the layers are generally unknown, a
quarter-wave plate (Q) set at 45$^{\circ}$ is used to convert the
signal photon to left circularly polarized light. A quarter-wave
plate is a polarization element, so that we must also include its
Jones matrix in the final expression for $\rm{\bf U}_1(\omega)$
(see Fig.~\ref{fig:QOCT-theory-setup}),
\begin{eqnarray}
  {\bf U}_1(\omega) &=& {\bf Q}(45)\,{\bf H}(\omega)\,{\bf Q}^{\dag}(45)
              \nonumber \\
        &=& \sum_{m=0}^{N}\,{\rm e}^{{\rm i}2\varphi^{(m)}}
                          {\bf Q}(45)\,{\bf B}^{(m)}\,{\bf r}_{m}\,{\tilde{\bf B}^{(m)}}\,{\bf
                          Q}^{\dag}(45),
\label{eq:sample_arm}
\end{eqnarray}
where the quarter-wave plate at 45$^\circ$ is defined as ${\bf
Q}(45) = {\rm e}^{{\rm
i}\frac{\pi}{4}\mbox{\boldmath$\sigma$}_{1}}$ and ${\bf
Q}^{\dag}(45) = {\bf Q}(-45)= {\rm e}^{-{\rm
i}\frac{\pi}{4}\mbox{\boldmath$\sigma$}_{1}}$.

The reference arm only contains a half-wave plate that can be set
at an angle $\theta$ to the horizontal axis (see
Fig.~\ref{fig:QOCT-setup}), so that $\rm{\bf U}_2$ can be written
as
\begin{equation}
   \rm{\bf U}_2(\omega') =  {\rm \bf R}(\theta) \, {\rm e}^{{\rm i}\frac{\pi}{2}\mbox{\boldmath$\sigma$}_{3}} \, {\rm \bf R}^\dag(\theta),
  \label{eq:ref_arm}
\end{equation}
where, for example, given a vertical input polarization, $2\theta
= 0^{\circ}$ selects vertical polarization and $2\theta =
90^{\circ}$ selects horizontal polarization.

We can now write the expression in Eq.~(\ref{eq:F}), assuming
frequency-independent isotropic reflection, i.e. ${\bf r}_{m} =
r_m \,\mbox{\boldmath$\sigma$}_{3}$, as
\begin{eqnarray}
  F(\omega) &=& {\bf e}_{\rm i}^{\dag}\,{\bf U}_2^{\dag}
                \sum_{m=0}^{N}\,{\rm e}^{{\rm i}2\varphi^{(m)}}
                          \,{\bf Q}(45)\,{\bf B}^{(m)}\,{\bf r}_{m}\,{\tilde{\bf B}^{(m)}}\,{\bf Q}^{\dag}(45)\,{\bf e}_{\rm s}
                \nonumber\\
            &=& {\bf e}_{\rm i}^\dag \, {\bf U}_2^\dag \, \sum_{m = 0}^N \,{\rm e}^{{\rm i}2\varphi^{(m)}} \, r_m \,{\bf u}_m(\omega)
    \label{eq:F-N-Layers}
\end{eqnarray}
where ${\bf u}_m(\omega) = {\bf Q}(45)\,{\bf
B}^{(m)}\,\mbox{\boldmath$\sigma$}_{3} \,{\tilde{\bf
B}^{(m)}}\,{\bf Q}^{\dag}(45)\,{\bf e}_{\rm s}$. For the sample
provided in Fig.~\ref{fig:N-layer-sample}, the general
Eqs.~(\ref{eq:lambda_zero})~and~(\ref{eq:lambda_tau}) become
\begin{equation}
  \Lambda_0 = \sum_{m=0}^{N}\sum_{n=0}^{N} r^{\ast}_{n}\,r_{m}\, \int d\Omega\,|\Phi(\Omega)|^{2}\,
            {\rm e}^{{\rm i}2[\varphi^{(m)}(\omega_0 + \Omega)-\varphi^{(n)}(\omega_0 + \Omega)]}\,
            {\bf u}_n^\dag(\omega_0 + \Omega)\,
            {\bf u}_m(\omega_0 + \Omega),
  \label{eq:self-int-term-general}
\end{equation}
and
\begin{equation}
  \Lambda({\tau}) = \sum_{m=0}^{N}\sum_{n=0}^{N} r^{\ast}_{n}\,r_{m}\, \int d\Omega\,|\Phi(\Omega)|^{2}\,
            {\rm e}^{{\rm i}2[\varphi^{(m)}(\omega_0 + \Omega)-\varphi^{(n)}(\omega_0 - \Omega)]}\,
            F_n^\ast(\omega_0 - \Omega)\,
            F_m(\omega_0 + \Omega)
            {\rm e}^{-{\rm i}\Omega\tau},
  \label{eq:cross-int-term-general}
\end{equation}
where $F_m(\omega) = {\bf e}_{\rm i}^{\dag}\,{\bf
U}_2^{\dag}\,{\bf u}_m(\omega)$. By substituting
Eqs.~(\ref{eq:self-int-term-general}) and
(\ref{eq:cross-int-term-general}) into Eq.~(\ref{eq:R-simple}), we
construct the final expression for the quantum interferogram.  In
the following sections, we investigate several special samples to
explain the features contained in the quantum interferogram and to
determine a method for extracting sample information.

\subsection{Single reflective layer}
If we consider the special case of a single isotropic reflector
buried under a birefringent layer of thickness $d_1 \equiv z_1$,
Eq.~(\ref{eq:N-layer-function}) can be written as
\begin{eqnarray}
        {\bf H}(\omega) &=& {\bf S}_{1}\,{\bf r}_{1}\,{\tilde{\bf S}}_{1}
        \nonumber\\
                        &=& {\rm e}^{{\rm i}\,2\Delta_{1}}\,{\bf B}_{1}\,{\bf r}_{1}\,{\tilde{\bf
                        B}_{1}}
  \label{eq:1-layer-sample}
\end{eqnarray}
whereupon Eq.~(\ref{eq:F}) becomes
\begin{eqnarray}
  F(\omega) &=& {\rm i}\,r_{1}\,{\rm e}^{{\rm i}\,2\Delta_{1}}\,\,({\bf e}_{i}^{\dag}\,{\bf U}_2^{\dag}\,
               \left[({\rm i}\,\sin\delta\,\sin 2\alpha_1)\,{\bf I} + (\cos\delta)\, \mbox{\boldmath$\sigma$}_{1} +
                              (\sin\delta\,\cos 2\alpha_1)\, \mbox{\boldmath$\sigma$}_{3} \right]\,{\bf e}_{s})
                \nonumber\\
            &=& {\rm i}\,r_{1}\,{\rm e}^{{\rm i}\,2\Delta_{1}}\,
               F_1(\omega),
    \label{eq:F-1-layer}
\end{eqnarray}
with
\begin{equation}
    F_1(\omega) = \cos\delta(\omega) \, \cos 2\theta + \sin\delta(\omega) \, \sin2\theta\,e^{2{\rm
    i}\alpha_1}.
    \label{eq:F1}
\end{equation}
We have made use of the properties of the Pauli spin matrices and
the fact that ${\bf e}_{\rm s} = \mbox{\boldmath$\sigma$}_{3}{\bf
e}_{\rm s}$ and ${\bf e}_{\rm i} =
\mbox{\boldmath$\sigma$}_{1}{\bf e}_{\rm s}$.

For a single reflector, Eqs.~(\ref{eq:self-int-term-general}) and
(\ref{eq:cross-int-term-general}) therefore become
\begin{equation}
  \Lambda_0 = |r_{1}|^2\, \int d\Omega\,|\Phi(\Omega)|^{2}\,
  \label{eq:self-int-term-1-layer}
\end{equation}
and
\begin{equation}
  \Lambda({\tau}) = |r_{1}|^2\, \int d\Omega\,|\Phi(\Omega)|^{2}\,
            {\rm e}^{{\rm i}2[\Delta_{1}(\omega_0 + \Omega)-\Delta_{1}(\omega_0 - \Omega)]}\,
            F_1(\omega_0 + \Omega)\,
            F_1^\ast(\omega_0 - \Omega)
            {\rm e}^{-{\rm i}\Omega\tau},
  \label{eq:cross-int-term-1-layer}
\end{equation}
respectively. The varying term can be further simplified if we
expand the propagation constant $\beta(\omega)$ in the expression
for the phase delay, $\Delta_1(\omega) = \omega \,{\bar n}\,z_1/c
= \beta(\omega)\,z_1$. The quantity $\beta(\omega_{0}+\Omega)$ is
expanded to second order in $\Omega$ so that
$\beta(\omega_{0}+\Omega)\approx\beta_{0}+\beta'\Omega+\frac{1}{2}\beta''\Omega^{2}$,
where $\beta'$ is the average inverse of the group velocities
$v_{\rm o}$ and $v_{\rm e}$ at $\omega_{0}$, and $\beta''$
represents the average group-velocity dispersion (GVD).  It is
clear that second-order dispersion is cancelled in the simplified
expression for the varying term, which is given by
\begin{equation}
  \Lambda({\tau}) = |r_{1}|^2\, \int d\Omega\,|\Phi(\Omega)|^{2}\,
            F_1(\omega_0 + \Omega)\,
            F_1^\ast(\omega_0 - \Omega)
            {\rm e}^{-{\rm i}\Omega(\tau + 4\beta'z_1)}.
  \label{eq:cross-int-term-1-layer-final}
\end{equation}

\begin{figure}
\includegraphics[clip,width=14cm]{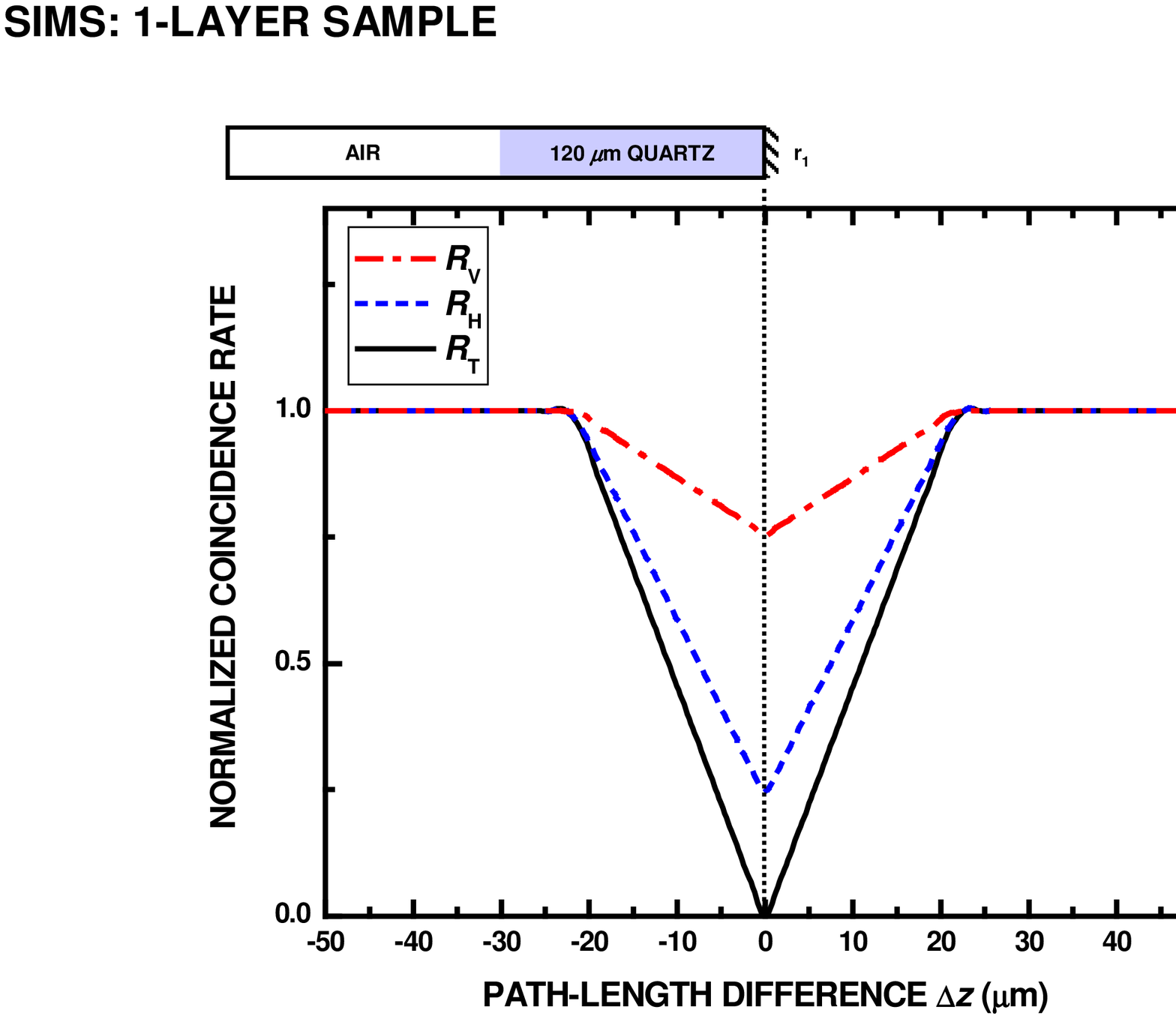}
\caption{Simulation results for a single reflector buried beneath
120~$\mu$m of quartz with $n_{\rm e} = 1.54661$, $n_{\rm o} =
1.53773$, and $|r_1|^2 = 1$, using the scheme shown in
Fig.~\ref{fig:QOCT-setup}. The optical axis of the quartz sample
is aligned with the horizontal axis in the laboratory frame so
that $\alpha_1 = 0$.  The top two curves represent the normalized
coincidence rate when the sample photon is mixed with a vertically
polarized ($R_{\rm V}$, dash-dot curve) or a horizontally
polarized ($R_{\rm H}$, dashed curve) reference photon. The solid
curve represents the total coincidence rate ($R_{\rm T}$) from
which the reflectance of the layer can be recovered.}
\label{fig:1-layer-sim}
\end{figure}

Figure~\ref{fig:1-layer-sim} displays the expected curves for a
single reflector buried beneath 120~$\mu$m of quartz with $n_{\rm
e} = 1.54661$, $n_{\rm o} = 1.53773$, and $|r_1|^2 = 1$, using the
scheme shown in Fig.~\ref{fig:QOCT-setup}. For this simulation, we
ignore the frequency dependence of $\delta$ [$\delta\equiv
\delta(\omega_{0})$], assume that $\int
d\Omega\,|\Phi(\Omega)|^{2} = 1$, and select the fast axis of the
quartz sample to be aligned with the horizontal axis in the
laboratory frame so that $\alpha_1 = 0$. The SPDC spectrum is
calculated explicitly via solutions to the phase-matching
conditions using published Sellmeier equations for BBO.  We are
interested in the particular case of degenerate, collinear type-II
phase matching.  The top two curves represent the expected
coincidence rate, normalized by $\Lambda_0$, when the sample
photon is mixed with a vertically polarized ($R_{\rm V}$, dash-dot
curve) or a horizontally polarized ($R_{\rm H}$, dashed curve)
photon from the reference arm. The solid curve represents the
re-normalized total coincidence rate [$R_{\rm T} = (R_{\rm V} +
R_{\rm H} - \Lambda_0)/\Lambda_0$] from which the reflectance of
the layer can be recovered.

The material properties are revealed by the relative values at the
center of the dip where $\tau = -4\beta'z_1$. At this value of
$\tau$, the path-length difference between the arms of the
interferometer is zero and there is maximal quantum interference.
If we neglect any frequency dependence in the birefringence, we
can substitute Eq.~(\ref{eq:F1}) into
Eq.~(\ref{eq:cross-int-term-1-layer-final}) and write an
expression for the coincidence rate at the center of the dip as
\begin{equation}
  \Lambda(\tau=-4\beta'z_1) = |F_1|^2 = |\cos\delta \, \cos2\theta + \sin\delta \,
\sin2\theta\,e^{2{\rm i}\alpha_1}|^2.
  \label{eq:1-layer-visibility}
\end{equation}
In the particular case when we select the linear-rotator angle
$2\theta$ to be either $0^\circ$ or $90^\circ$, corresponding to a
polarization of the reference photon that is horizontal or
vertical, respectively, we obtain
\begin{eqnarray}
    \Lambda_{\rm H} &=& \cos^2 \delta \nonumber \\
    \Lambda_{\rm V} &=& \sin^2 \delta.
\end{eqnarray}
It is possible to determine the value of $\delta$, or the
birefringence $\Delta n$, by forming a ratio of these rates at
$\Delta z = 0$:
\begin{equation}\label{eq:delta}
   \delta = \tan^{-1} \left[{\frac{\Lambda_{\rm V}}{\Lambda_{\rm H}}}
   \right]^{\frac{1}{2}} = \omega_0\,\Delta
   n\,z_1 /c.
\end{equation}

We can neglect the frequency dependence of $\delta(\omega)=
\omega\Delta n(\omega)\,z_1 /c = \delta\beta(\omega)z_1$ when
$\delta^{\prime}(\omega_0)\,\Delta\Omega \ll 1$, where
$\Delta\Omega$ is the bandwidth of the SPDC spectrum.  In this
limit, the width of the interference dip is larger than the delay
between the signal and idler fields resulting from the
birefringence of the layer.  If the bandwidth of the SPDC spectrum
is increased, the opposite limit can be realized, namely
$\delta^{\prime}(\omega_0)\,\Delta\Omega \gg 1$.  In this case,
the interference pattern comprises of three regions: the expected
central dip at $\tau = -4\beta'z_1$ provides the value of
$\delta(\omega_0) \equiv \delta$ as in Eq.~(\ref{eq:delta}); and
two additional satellite interference patterns centered at $\tau =
-[4\beta' \pm 2\delta\beta']z_1$, where $\delta\beta'$ is the
coefficient of the first-order expansion of $\delta\beta$ in
$\Omega$, and provide information about the group-velocity
dispersion in the layer.

Since we choose the linear-rotator angles $2\theta$ to be either
$0^\circ$ or $90^\circ$, any dependency of the coincidence rate
according to the orientation angle $\alpha_1$ is lost.  It is
possible, however, to extract the value of $\alpha_1$ by using a
technique that is analogous to null ellipsometry. In the reference
arm, if we combine the linear-rotator used to rotate the linear
input polarization state ${\bf e}_{\rm i}$ with a quarter-wave
plate to transform the linear polarization into a general
elliptical state, it is possible to exactly match any polarization
state in the sample arm.  This transformation is given by
\begin{equation}
   {\bf U}_{2}\,{\bf e}_{\rm i} = \frac{1}{\sqrt{2}}
  \begin{bmatrix}
    \cos 2\theta + {\rm i}\cos2(\phi - \theta) \\
    \sin 2\theta + {\rm i}\sin2(\phi - \theta)
  \end{bmatrix},
   \label{eq:ref-pol-transform}
\end{equation}
where $\phi$ is the angle of the quarter-wave plate and $\theta$
is the angle of the linear rotator fast axes with respect to the
horizontal axis. In the special case when $\phi = 2\theta$, we
revert to the case of a single linear rotator as in our previous
example.

When the polarization in the reference arm is selected by this
cascade of polarization elements, we can write Eq.~(\ref{eq:F1})
as
\begin{equation}
    F_1(\omega) = \left\{
                         \cos\delta(\omega)\,[\cos2\theta - {\rm i}\cos2(\phi - \theta)] +
                         \sin\delta(\omega)\,[\sin 2\theta - {\rm i}\sin2(\phi - \theta)]\,
                         e^{2{\rm i}\alpha_1}
                 \right\}.
    \label{eq:F1-nulling}
\end{equation}
If the values of $\phi$ and $\theta$ are adjusted so that the
polarization state in the reference arm is exactly orthogonal to
that in the sample arm, $|F_1|^2 = 0$ and the coincidence count
rate will be maximized. The value for $\alpha$ can then be
determined by solving the following conditions of orthonormality,
namely, the real and/or imaginary parts of
Eq.~(\ref{eq:F1-nulling}) must equal zero
\begin{eqnarray}
     \cos\delta\,\cos2\theta + \sin\delta\,\sin
     2\theta\,\cos2\alpha_1 - \sin\delta\,\sin2(\phi -
     \theta)\,\sin2\alpha_1 &=& 0\\ \nonumber
     -\cos\delta\,\cos2(\phi - \theta) + \sin\delta\,\sin
     2\theta\,\sin2\alpha_1 - \sin\delta\,\sin2(\phi -
     \theta)\,\cos2\alpha_1 &=& 0.
    \label{eq:F1-nulling-max-condition}
\end{eqnarray}
If the value of $\delta$ is known, then only one of these
equations is required.

\subsection{Two reflective layers}
A sample with reflections from two surfaces separated by a
birefringent material can be expressed as
\begin{eqnarray}
  {\bf H}(\omega) &=& {\bf r}_{0} + {\bf S}_{1}\,{\bf r}_{1}\,{\tilde{\bf
  S}_{1}}
                \nonumber\\
                  &=& {\bf r}_{0} + {\rm e}^{{\rm i}\,2\Delta_{1}}\,{\bf
B}_{1}\,{\bf r}_{1}\,{\tilde{\bf B}_{1}}
    \label{eq:2-layer-sample}
\end{eqnarray}
where the subscripts $0$ and $1$ denote the first and second
boundaries, respectively.  In this case, the function in
Eq.~(\ref{eq:F}) becomes
\begin{equation}
  F(\omega) =  {\rm i}\,r_0\,F_0 + {\rm i}\,r_1\,{\rm e}^{{\rm i}\,2\Delta_{1}}\,
  F_1,
\end{equation}
where $F_0 = \cos2\theta$ and $F_1$ has been provided in
Eq.~(\ref{eq:F1}).

For two reflectors separated by a birefringent medium, the
constant and varying contributions from
Eqs.~(\ref{eq:self-int-term-general}) and
(\ref{eq:cross-int-term-general}) become
\begin{equation}
  \Lambda_0 = |r_0|^2 \, \int d\Omega\,|\Phi(\Omega)|^{2} +
              |r_1|^2 \, \int d\Omega\,|\Phi(\Omega)|^{2} +
               r_0^\ast \,r_1 \, {\rm e}^{{\rm i}2\beta_0z_1}\,\int d\Omega\,|\Phi(\Omega)|^{2}
               {\bf u}^\dag_0 \,{\bf u}_1(\omega_0 + \Omega)\,{\rm e}^{{\rm i}2(\beta'\Omega + \beta''\Omega^2)z_1} + cc.
  \label{eq:self-int-term-2-layer}
\end{equation}
and
\begin{equation}
  \Lambda({\tau}) = |r_0|^2 \, g^{(0)}(\tau) +
                    |r_1|^2 \, g^{(1)}(\tau - 4\beta'z_1) +
                     r_0^\ast\,r_1 \, g^{(01)}_{\rm d}(\tau - 2\beta'z_1) \, {\rm e}^{{\rm i}2\beta_0z_1} +
                     cc.,
  \label{eq:cross-int-term-2-layer}
\end{equation}
respectively, where the subscript ${\rm d}$ denotes a contribution
that is subject to even-order dispersion and $cc$ indicates the
complex conjugate, with
\begin{equation}
g^{(m)}(\tau) = \int d\Omega\,|\Phi(\Omega)|^{2}\, F_m(\omega_0 +
\Omega)\,F_m^\ast(\omega_0 - \Omega)\,{\rm e}^{-{\rm i}\Omega\tau}
\nonumber
\end{equation}
and
\begin{equation}
g_{\rm d}^{(mn)}(\tau) = \int d\Omega\,|\Phi(\Omega)|^{2}\,
F_m(\omega_0 + \Omega)\,F_n^\ast(\omega_0 - \Omega)\,{\rm e}^{{\rm
i} 2 z_1 \beta'' \Omega^2}\,{\rm e}^{-{\rm i}\Omega\tau}.
\nonumber
\end{equation}
The first two terms in Eq.~(\ref{eq:self-int-term-2-layer}) are
contributions to the constant coincidence rate arising from each
of the two interfaces in the material.  The third term introduces
a contribution only when these interfaces have a separation that
is less than the coherence length of the signal photon.  The first
two terms in Eq.~(\ref{eq:cross-int-term-2-layer}) represent dips
arising from reflections from the first and second surfaces. The
third term, which appears midway between these dips, arises from
the interference between probability amplitudes associated with
each of these reflections.

\begin{figure}
\includegraphics[clip,width=14cm]{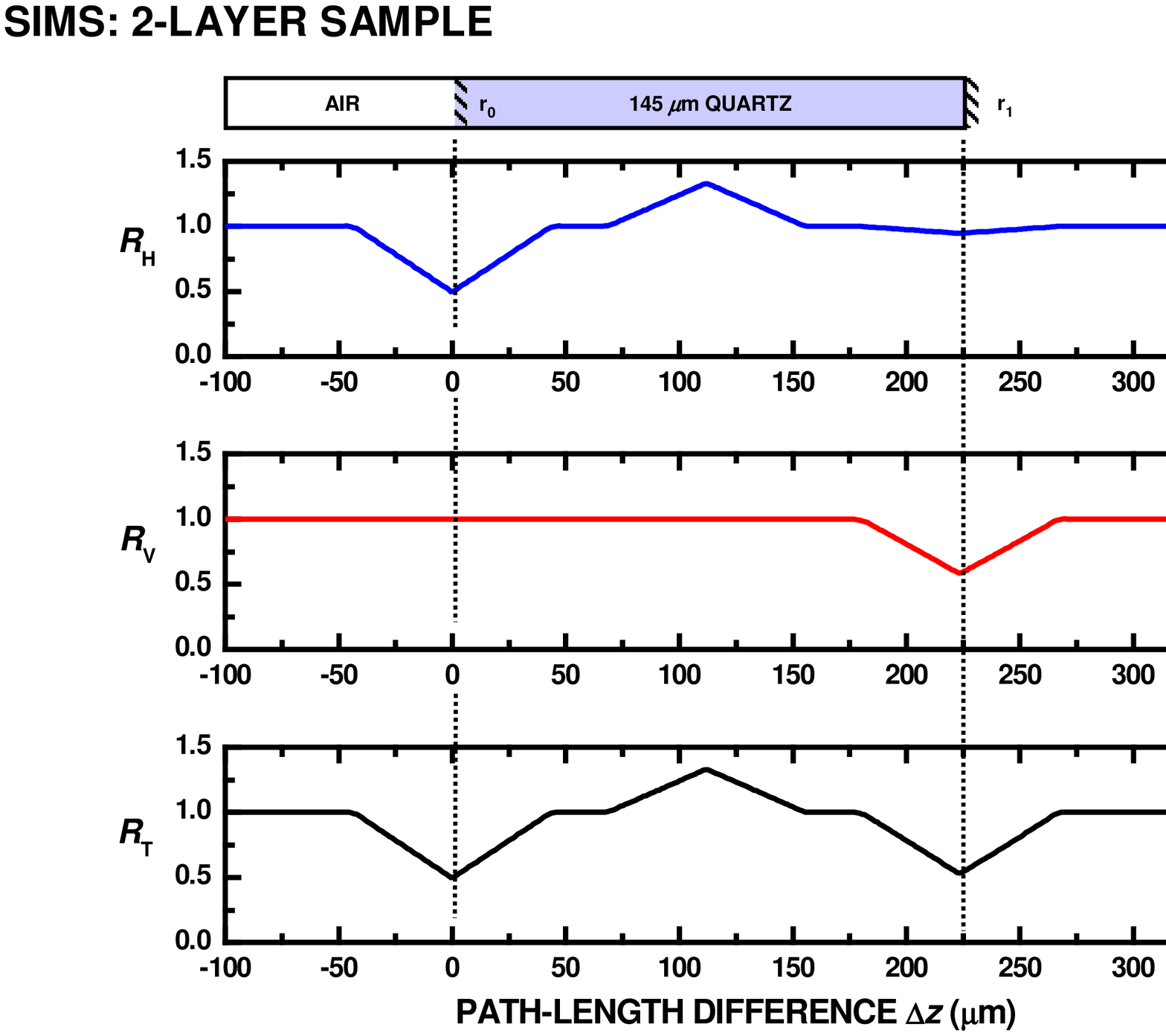}
\caption{Simulation results for a 145-$\mu$m quartz sample with
reflections from each interface using the scheme shown in
Fig.~\ref{fig:QOCT-setup}.  For this calculation, $n_{\rm e} =
1.54661$, $n_{\rm o} = 1.53773$, and $|r_0|^2 = |r_1|^2$.  The
optical axis of the quartz is aligned with the horizontal axis in
the laboratory frame so that $\alpha_1 = 0$. The top two plots
represent the normalized coincidence rate when the sample photon
is mixed with a horizontally polarized ($R_{\rm H}$) or a
vertically polarized ($R_{\rm V}$) reference photon. The bottom
trace represents the total coincidence rate ($R_{\rm T}$) from
which the relative reflectance of each interface can be recovered.
The separation of the dips is given by the optical path length of
the quartz ${\bar n}L \simeq 224~\mu$m.} \label{fig:2-layer-sim}
\end{figure}

Figure~\ref{fig:2-layer-sim} provides numerical results for a
145-$\mu$m quartz sample with reflections from each of the two
surfaces. For this calculation, again $n_{\rm e} = 1.54661$,
$n_{\rm o} = 1.53773$, we ignore the frequency dependence of
$\delta$, and the fast axis of the quartz plate is aligned with
the horizontal axis in the laboratory frame so that $\alpha_1 =
0$. The magnitude of the reflectance from each surface is assumed
to be the same so that $|r_0|^2 = |r_1|^2$. The SPDC spectrum is
calculated explicitly via solutions to the phase-matching
conditions using published Sellmeier equations for BBO.  We are
interested in the particular case of degenerate, collinear type-II
phase matching. The top two plots represent the expected rate of
coincidence when the sample photon is mixed with a horizontally
polarized ($R_{\rm H}$) or a vertically polarized ($R_{\rm V}$)
reference photon. The bottom trace represents the re-normalized
total coincidence rate [$R_{\rm T} = (R_{\rm V} + R_{\rm H} -
\Lambda_0)/\Lambda_0$] from which the relative reflectance and
positions of each interface can be determined.

In the $R_{\rm V}$ curve (middle trace), there is no dip at the
first interface since the polarization mode reflected from this
interface is solely horizontal.  The polarization state is altered
via propagation through the birefringent material and contains
both vertically and horizontally polarized photons at reflection
from the second interface. The peak between the two interfaces in
$R_{\rm H}$ (top trace) and $R_{\rm T}$ (bottom trace) is result
of interference between each layer. This peak (which can
alternatively become a dip depending on the phase accumulated
between the layers) is susceptible to dispersion in the sample,
unlike the dips that correspond to sample layers.  Thus the
dispersion properties of the material can be extracted from this
feature.

In summary, we ascertain that three experiments are required to
completely determine the sample properties.  We first select the
reference arm polarization to be horizontal (H) and measure the
quantum interferogram $R_{\rm H}$ by recording the coincidence
rate of photons arriving at the two detectors as the path-length
delay $c\tau$ is scanned. The reference arm polarization is then
rotated into the orthogonal vertical (V) polarization and a second
measurement is made to measure the quantum interferogram $R_{\rm
V}$. The third measurement is made by selecting a value of $c\tau$
that coincides with the position of a layer.  The angles of the
polarization elements in the reference arm are then adjusted to
maximize the coincidence rate.

The sample properties are found as follows: by forming a ratio of
$\Lambda_{\rm V}$ and $\Lambda_{\rm H}$ at a value of $c\tau$ that
coincides with the position of a layer, we can determine the value
of the birefringence contained in the parameter $\delta$; using
the angles from the polarization elements in the reference arm,
$\alpha$ can be found from solving the equations for
orthonormality.  This technique is similar to nulling techniques
in ellipsometry; and the total quantum interferogram $R_{\rm T}$
can be computed from the sum of $R_{\rm H}$ and $R_{\rm V}$, then
readjusted for the dc offset given by the constant term
$\Lambda_0$, i.e. $R_{\rm T} = (R_{\rm V} + R_{\rm H} -
\Lambda_0)/\Lambda_0$. The $R_{\rm T}$ curve provides the
path-length delay between the interfaces as well as the ratio of
the relative reflectance from each layer.

\section{CONCLUSION}
We have set forth a new polarization-sensitive QOCT (PS-QOCT)
scheme and provide a general Jones matrix theory for analyzing its
operation. PS-QOCT provides a means for determining information
about the optical path length between isotropic reflectors, the
relative magnitude of the reflectance from each interface, the
birefringence of the material between the interfaces, and the
orientation of the optical axis $\alpha$ of the sample. Inasmuch
as PS-QOCT is immune to sample dispersion, measurements are
permitted at depths greater than those accessible via ordinary
optical coherence tomography.

\begin{acknowledgments}
This work was supported by the National Science Foundation; the
Center for Subsurface Sensing and Imaging Systems (CenSSIS), an
NSF Engineering Research Center; and the David and Lucile Packard
Foundation.
\end{acknowledgments}

\bibliographystyle{apsrev}
\bibliography{qil-bib-1-abbrev}

\clearpage

\clearpage

\clearpage

\clearpage

\clearpage

\end{document}